\documentclass[10pt,journal,compsoc]{IEEEtran}

\usepackage[nocompress]{cite}
\usepackage{amsmath,amssymb,amsfonts}
\usepackage{algorithmic}
\usepackage[pdftex]{graphicx}
\usepackage{textcomp}
\usepackage{epsfig} 
\usepackage{epstopdf} 
\usepackage{mathptmx}
\usepackage{times}
\usepackage{epstopdf}
\usepackage{setspace}
\usepackage{float}
\usepackage{booktabs}
\usepackage[center]{caption}
\usepackage{multicol}
\usepackage{dblfloatfix}

\begin{document}

\title{A Stochastic Optimal Control Model with Internal Feedback and Velocity Tracking for Saccades}

\author{Varsha~V.,~Aditya~Murthy, and~Radhakant~Padhi
\IEEEcompsocitemizethanks{\IEEEcompsocthanksitem Varsha V was with the Centre for Biosystems Science and Engineering, Indian Institute of Science, Bangalore, India 560012, E-mail: varshav@iisc.ac.in.\protect\\
\IEEEcompsocthanksitem A. Murthy is with Centre for Neuroscience and Centre for Biosystems Science and Engineering, Indian Institute of Science, Bangalore, India 560012, E-mail: adi@iisc.ac.in. \protect\\
\IEEEcompsocthanksitem R. Padhi is with Department of Aerospace Engineering and Centre for Biosystems Science and Engineering, Indian Institute of Science, Bangalore, India 560012, E-mail: padhi@iisc.ac.in}
}

\IEEEtitleabstractindextext{%
\begin{abstract}
A stochastic optimal control based model with velocity tracking and internal feedback for saccadic eye movements is presented in this paper. Recent evidence from neurophysiological studies of superior colliculus suggests the presence of a dynamic input to the saccade generation system that encodes saccade velocity, rather than just the saccade amplitude and direction. The new evidence makes it imperative to test if saccade control can use a desired velocity input which is the basis for the proposed velocity tracking model. The model is validated using behavioral data of saccades generated by healthy human subjects. It generates trajectories of horizontal saccades made to different amplitudes as well as predicts vertical and oblique saccade behavior. This paper presents the first-ever model of the saccadic system in an optimal control framework using an alternate interpretation of velocity-based control, contrary to the dominant end-point based models available in the literature.
\end{abstract}

\begin{IEEEkeywords}
Saccadic eye movement, Internal feedback, Velocity tracking in saccades, Stochastic optimal control.
\end{IEEEkeywords}}

\maketitle




\IEEEraisesectionheading{\section{Introduction}\label{sec:intro}}

\IEEEPARstart{S}{accades} are very fast and accurate movements. Since they are rapid, saccades cannot make use of sensory feedback information about the state of the eye for its precise execution~\cite{guthrie1983corollary, wurtz2008neuronal}. So they are thought to be largely pre-programmed to achieve the goal of landing the eye on the target. The idea of pre-programmed or feed-forward control has been used in the framework of optimal control theory to understand the basis of normative saccades. Different models were suggested with cost functions involving minimum time, minimum jerk, minimum energy, and their combinations to explain saccade characteristics like the main sequence (which is the relationship between amplitude, peak velocity, and the duration of saccades). All these models have considered the saccadic system to be deterministic even though it is well established that noise is an inherent feature of the motor system \cite{faisal2008noise,smeets2003nature,van2007sources}. Hence, models that do not incorporate stochasticity are incomplete.

\par More recent optimal control models have however incorporated the presence of noise. In such stochastic models, the noise is typically considered to be signal-dependent and this noise structure is motivated by neurophysiological evidence provided in \cite{jones2002sources}. The study in \cite{harris2006main} proposed a speed-accuracy trade-off (SAT) model of saccade control in the stochastic optimal control framework. The model interpreted that the saccadic system tries to minimize a cost that depends on the variance in displacement at the end of the saccade and the time taken for the saccade. Signal dependent noise was incorporated in the model as an extra additive signal acting on the optimal control signal. This model was able to capture the main sequence saccade characteristics well and suggested that the saccade kinematics are generated such that an optimal trade-off between its speed and accuracy is achieved.

\begin{figure*}
\centering
\includegraphics[width=0.8\textwidth,height=0.25\textwidth]{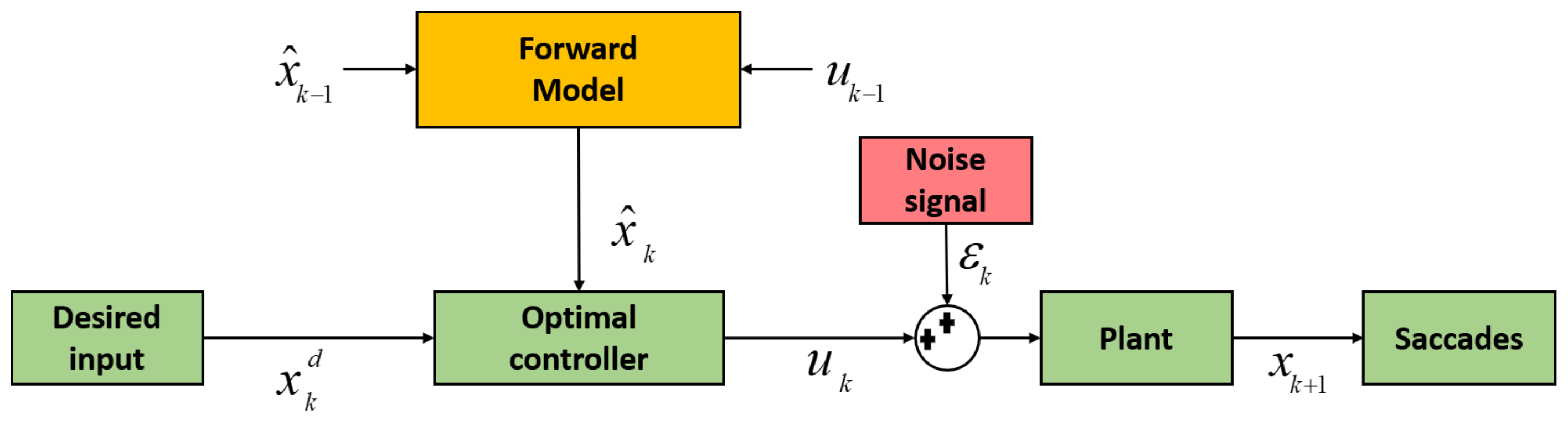}
\caption{Block diagram of velocity tracking model}
\label{fig:modelschema}
\end{figure*}

\par However, the pre-planning of a feed-forward control action for all time points in the future in a goal-directed eye movement may not be desirable and productive in the presence of stochasticity. In such a scenario, a moment-to-moment control through sensory feedback becomes important. The saccadic system is devoid of such a feedback control due to the rapid time course that precludes any sensory feedback information from being relevant. However, there is much evidence suggesting the use of non-sensory feedback, also known as an internal feedback mechanism. The earliest evidence of internal feedback affecting eye movements was suggested through perturbation experiments in monkeys \cite{keller1996endpoint}. They found that although saccades were interrupted mid-flight by stimulation and target information was made unavailable, the interrupted saccade foveated the target accurately on the removal of the interrupting stimulation. The existence of such internal feedback has also been confirmed by other studies using different saccade tasks \cite{richardson2011time,xu2011tms}. These studies have shown online correction of saccade trajectories, suggesting the presence of internal feedback-based control of saccades. Further, a transcranial magnetic stimulation (TMS) study showed that a TMS perturbed saccade trajectory was corrected within the same saccade to produce accurate movements to the target without relying on visual input. The study claimed that the oculomotor commands are monitored as they unfold via an internal feedback mechanism and is corrected for the perturbation.

\par The ability to monitor and control saccadic motor commands in real-time using internal feedback was incorporated using an extension of the optimal control theory approach called the optimal feedback control theory \cite{todorov2002optimal,todorov2005stochastic}. The theory proposed that the control signal is generated at each time step based on the estimate of the current state of the system, using a forward model present in the internal feedback loop. Since eye movements cannot make use of sensory feedback, it is hypothesized that the real-time estimate of the state is obtained through the forward model, using a copy of the control signal given to the muscles. In \cite{chen2008adaptive}, it was shown that such an optimal feedback model with a cost function that penalizes control effort, duration of movement, and the error in the state from the desired final state works well in explaining adaptation in saccade behavior. The model was posed as a regulator problem where the system was trying to regulate the states at the final saccadic displacement to be achieved by the eye and zero final velocity. The model has also been shown to explain normal saccade trajectories in \cite{shadmehr2012biological}.

\par In both the feed-forward \cite{harris2006main} and feedback model of saccade control \cite{chen2008adaptive}, the goal is to achieve a desired final displacement (i.e. saccade amplitude). This is based on the observations made by neurophysiological studies about the presence of a spatial map in the midbrain superior colliculus (SC), which is thought to be the center providing the desired goal to the brainstem saccade generation circuit. The microstimulation of different locations on this map generated saccades with a particular direction and amplitude. Hence, modeling studies have suggested that the desired endpoint or final displacement acts as the input to the saccade generation system (for instance see \cite{robinson1973models, Scudder1988}). But reversible inactivation of SC has shown that the trajectory of the saccade itself is altered apart from the amplitude and direction of the saccades\cite{aizawa1998reversible}. There are also a few studies like \cite{sparks1990population, berthoz1986some} and \cite{reppert2018neural} which have shown the average firing rates in collicular cells to be correlated with peak saccadic velocity. However, since saccades show a tight nonlinear coupling between amplitudes and peak velocities, the saccade generation system cannot achieve a particular saccade amplitude based on just the peak velocity information. Rather the SC neurons would require to encode the entire velocity profile, as its integral would give the amplitude. Thus, suggesting that the saccade kinematics could be dynamically encoded by SC neurons. This hypothesis is supported by another recent neurophysiological study that showed a robust correlation of instantaneous activity in superior colliculus to the instantaneous eye velocity throughout the saccade duration \cite{smalianchuk2018instantaneous}. All this evidence has led to the strengthening of the idea that the desired signal input to the saccade generation system, encoded by the SC neurons, could be in the form of velocity information.

\par Despite the evidence, there are no optimal control models of saccades that are designed to use the desired velocity signal as input to the brainstem saccade generation system. Hence in this work, an optimal control model of the saccadic system that tracks a desired velocity has been developed and the ability of such a control mechanism to predict saccade behavior has been tested. The presence of both internal feedback and stochasticity in the saccadic system is incorporated into the model. The proposed stochastic model can generate the control signal that allows tracking of the desired optimal velocity even in the presence of noise. The model is capable of explaining the behavioral data of saccades collected from human subjects. It captures the main sequence relationship across multiple amplitudes of saccades and also predicts the trajectories of saccades made in different directions.\\

\section{Proposed Velocity Tracking Saccade Model} \label{sec:model}

\par A velocity tracking stochastic optimal control model with internal feedback is proposed for saccade generation. The goal of the model is to track a desired saccade velocity. The average behavior across repetitions is assumed to be the optimal human behavior and hence the desired velocity input for each individual is modeled as the mean velocity of all saccades produced by the individual to each target during the experiments. An optimal control signal is designed for tracking the desired velocity using an estimate of the state from a forward model (described in Section \ref{sec:fwd_mdl}). The control signal is assumed to be corrupted by signal-dependent additive noise and its characteristics are discussed in Section \ref{sec:noise}. The noisy control signal is the input to the oculomotor plant that generates the saccade trajectories (described in Section \ref{sec:plant}). A schematic block diagram of the model is given in Fig. \ref{fig:modelschema} and the details about the different blocks constituting the model are described in the following subsections.
\subsection{Oculomotor Plant Dynamics}\label{sec:plant}

\par The oculomotor plant is modeled as a lumped system consisting of the extraocular muscles and eyeball together and is represented by the block labeled as the plant in Fig. \ref{fig:modelschema}. A second-order spring-mass-damper model which provides a good approximation of this system has been used to derive the plant dynamics using a simple force balance equation \cite{robinson1981use}. The input to the dynamic system is the firing rate produced at the motor neurons innervating the oculomotor muscle and it is assumed to be equivalent to the torque required to move the eye. If the input firing rate is represented by $R(t)$, the force balance equation can be written as
\begin{equation}
R(t)=J\ddot{\theta}(t)+B\dot{\theta}(t)+K\theta(t)
\end{equation}
where, the moment of inertia, viscous coefficient, and elastic coefficient of the eyeball system is given by $J$, $B$, and $K$ respectively. 
This equation can be represented as a continuous linear dynamical system in state-space form as
\begin{equation}
\dot{x}(t)=A_cx(t)+B_cu(t)\label{eqn: 2a}
\end{equation}
where $u(t)$ is the control signal defined as equivalent to the firing rate $R(t)$. The state $x(t)\triangleq\left[\theta(t), \dot{\theta}(t)\right]^T$ and state matrices $A_c$ and $B_c$ are given by
\begin{align}
\label{eqn: 2}
A_c\triangleq& \begin{bmatrix} 
0 & 1 \\
\frac{1}{ \tau_{1}\tau_{2} } & -(\frac{\tau_1+\tau_2}{\tau_1\tau_2})\\
\end{bmatrix}\nonumber\\
B_c\triangleq& \begin{bmatrix} 
0 \\
\frac{1}{\tau_{1}\tau_{2}}\\
\end{bmatrix}
\end{align}
In (\ref{eqn: 2}), $\tau_{1}$ and $\tau_{2}$ are given by $\tau_{1}=B/K\text{ and }\tau_{2}=J/B$, with values assumed to be $223$ ms and $14$ ms, respectively as in \cite{robinson1981use, harris1998signal}. This system dynamics is valid only when considering horizontal saccades ($0^\circ$ or $180^\circ$ directions). 
\par For oblique saccades (see Section \ref{sec:design}), the system dynamics involves a horizontal component and a vertical component separately. Hence, in the case of oblique saccades the state $x(t)\triangleq\left[\theta^h(t),\dot{\theta }^h(t),\theta^v(t),\dot{\theta }^v(t)\right]^T$ and control $u(t)\triangleq\left[u^h(t), u^v(t)\right]^T$. The state matrices ${{A}_{c}}\text{ and }{{B}_{c}}$ are defined as follows
\begin{align}
\label{eqn: obldyn}
A_c\triangleq& \left[\begin{matrix}
0 & 1 & 0 & 0\\
\frac{-1}{\tau_1\tau_2} & -(\frac{\tau_1+\tau_2}{\tau_1\tau_2}) & 0 & 0 \\
0 & 0 & 0 & 1 \\
0 & 0 & \frac{-1}{\tau_1\tau_2} & -(\frac{\tau_1+\tau_2}{\tau_1\tau_2})\\ \end{matrix} \right] \nonumber\\ 
B_c\triangleq& {{\left[ \begin{matrix} 
0 & \frac{1}{{{\tau }_{1}}{{\tau }_{2}}} & 0 & 0\\
0 & 0 & 0 & \frac{1}{{{\tau }_{1}}{{\tau }_{2}}}\\ \end{matrix} \right]}}^{T}
\end{align}
This study has assumed that there is no coupling between the horizontal ad vertical components in the oculomotor plant dynamics, similar to \cite{chen2008adaptive}. This makes the oblique saccade formulation equivalent to solving two one dimensional problems of horizontal component and vertical component of the saccade separately. In reality, it is possible that there is coupling between components in case of oblique saccades. A more nuanced model of oculomotor plant would be needed for studying these coupling effects.
\par Subsequently, discrete optimal control is used to solve this problem and hence there is a need for discretization of the state dynamics given in (\ref{eqn: 2a}). The state dynamics in discrete form can be written for both the horizontal and oblique case as
\begin{equation}\label{eqn:discrete}
{{x}_{k+1}}={{A}}{{x}_{k}}+{{B}}{{u}_{k}}
\end{equation}
with, ${{A}}={{e}^{{{A}_{c}}\Delta t}}$ and ${{B}}=A_{c}^{-1}({{e}^{{{A}_{c}}\Delta t}}-I){{B}_{c}}$ representing the system matrices in discrete-time and $k$ denoting the discrete time-steps. A $\Delta t= 4ms$ is used for simulations, as we found that decreasing the time steps further did not have any influence on the simulation results. Note that one can also use the first order approximation in discretization which gives $A=A_c\Delta t+I$ and $B=B_c\Delta t$. However, in both cases the results had no appreciable difference.

\subsection{Noise in the control signal}\label{sec:noise}

\par Noise affects the saccadic system at different levels like target localization, movement planning, and execution \cite{van2007sources}. Here, it is assumed that the effect of all these noises would be reflected as an uncertainty added to the control signal acting on the plant. Interestingly, it turns out that the noise affecting the neural command is signal-dependent, which means that \textit{the standard deviation of the noise distribution scales with the mean value of the signal} and hence is dependent on the mean control signal itself \cite{harris1998signal}. Such a noise structure is motivated by previous literature that suggested that the signal-dependent noise in the control signal can capture saccade trajectories and the main sequence \cite{harris2006main}. There is also experimental evidence that suggests that the source of the signal-dependent noise in the case of isometric force production can be attributed to central voluntary control, rather than peripheral muscle components \cite{jones2002sources}. This suggests that the noise is added to the control signal $u_k$ before acting on the plant and the noise influence matrix in the system dynamics can be taken to be the same as the control influence matrix. Based on these observations, the stochastic system dynamics can be written as
\begin{equation}
\label{eqn:stocsd}
x_{k+1}=Ax_k+B (u_k+\epsilon_k),\quad{k=1,2,\ldots,n}
\end{equation} 
where, $n$ is the number of time-steps in the simulation and $\epsilon_k=\alpha u_kw_k$, where, $w_k$ represents a random variable drawn from zero-mean Gaussian distribution with a variance of one and $\alpha$ is the noise scaling factor in the signal-dependent noise definition\cite{harris1998signal}. Note that, $\alpha$ is a free parameter estimated from experimental data. This structure of the noise is very important in further analyses.
\subsection{Forward Model}\label{sec:fwd_mdl}

An important component in the proposed model is the presence of an internal feedback loop, which uses a copy of the control signal to estimate the state with the help of a forward model block (see Fig. \ref{fig:modelschema}). In a generic case, the estimate of the state would be obtained by combining the prediction in the current step with sensory observations obtained from output feedback. But in the case of the saccadic system, the output feedback is not available and hence the estimate at the current step depends only on the estimate in the previous step. Thus, given the copy of the un-corrupted control signal $u_k$, the forward model estimates the states at the next instant of time, $\hat{x}_{k+1}$ as follows
\begin{equation}\label{eqn:FWD}
\hat{x}_{k+1}=A_f\hat{x}_{k}+B_fu_k
\end{equation} 
where, $\hat{x}_{k}=\mathop{\mathbb{E}}[x_k]$. Since the forward model is assumed to be not affected by noise and can perfectly predict the states of the system, we take $A_f=A$ and $B_f=B$ respectively.\\

\section{Velocity Tracking Optimal Controller}
\par To investigate if the saccadic system utilizes a velocity tracking strategy to control saccades, an optimal control policy that is based on a desired velocity signal has to be calculated. The policy needs to convert the desired signal $x^d$ and the estimate of the state $\hat{x}$ at each step into a motor commands or control signals $u$. This transformation is carried out by the optimal controller block in Fig. \ref{fig:modelschema}. The details of finding a sequence of optimal control signals at all steps are discussed in the following subsections.

\subsection{Control formulation}
\label{sec:OC}
\par The cost function $V$ that is minimized during the movement (with N steps) is taken as the expectation of sum of costs at all steps from 1 to N and is given by 
\begin{equation}
\label{eqn:gen_cost}
V=\mathop{\mathbb{E}}\left[\sum_{i=1}^{N}\psi_i\right]
\end{equation}
where, $\psi_i$ is the cost at each step $i$ and $\mathop{\mathbb{E}}$ represents expectation operator. The expectation is taken because the state dynamics is stochastic and only its expectation can be known deterministically in this problem. The cost function is subject to the stochastic system dynamics $x_{k+1}=Ax_k+B (u_k+\epsilon_k)$ as given by (\ref{eqn:stocsd}), with the state vector ${{x}_{i}}={{\left[ {{\theta }_{i}},{{{\dot{\theta }}}_{i}} \right]}^{T}}$ consisting of angular displacement $\theta_i$ and angular velocity $\dot{\theta_i}$. The initial condition of the state is fixed and taken as the value of mean angular displacement $\theta^\mu_1$ and mean angular velocity $\dot{\theta}^\mu_1$ at the first time step in the experimental data. The final conditions of the states are free. There are no constraints involved in this problem.
\par The cost at each step is given by
\begin{equation}\label{eqn:general_stepcost}
\psi_i={{\left( {{x}_{i}}-x_{i}^{d} \right)}^{T}}{{Q}_{i}}\left( {{x}_{i}}-x_{i}^{d} \right)+u_{i}^{T}{{R}_{i}}{{u}_{i}},
\end{equation}
where, the desired state is $x_{i}^{d}={{\left[ \theta _{i}^{d},\dot{\theta }_{i}^{d} \right]}^{T}}$. The desired values of the state are modeled as the mean of the experimental data obtained for each subject across the trials. The cost at each step $i$ is a combination of two terms. The first term makes sure that the error between the desired system states and the achieved state is minimum. The second term minimizes the effort by penalizing the control signal magnitude. The error weightage matrix ${{Q}_{i}}$ and the control weightage matrix ${{R}_{i}}$ decides the distribution of the cost between the two terms. In this problem, ${{Q}_{i}}=[0,0\text{ ; }0,q]$ where $q$ is a free parameter. The structure of $Q_i$ was chosen to have weightage to only velocity error as only desired velocity was available at the input. The control weighting parameter ${{R}_{i}}$ is fixed to one. It is observed that this assumption did not affect the solution since this is equivalent to redefining original error weightage with a scaling factor of $1/R$ and the free parameter $q$ in the error weightage matrix absorbed this scaling effect.
\subsection{Control solution using dynamic programming approach}\label{sec:OC_D}
\par The optimal control problem is solved using dynamic programming approach \cite{shadmehr2012biological,chen2008adaptive}. The cost from any time step $k$ can be written based on (\ref{eqn:gen_cost}) as sum of costs from step $k$ to final step $N$ and is given by
\begin{equation}
\label{eqn:k_cost}
V_k=\mathop{\mathbb{E}}\left[\sum_{i=k}^{N}\psi_i\right]
\end{equation}
which can be then rewritten as
\begin{equation}
\label{eqn:uti_cost}
V_k=\mathop{\mathbb{E}}\left[\psi_k+V_{k+1}]=\mathop{\mathbb{E}}\left[\psi_k\right]+\mathop{\mathbb{E}}[V_{k+1}\right]
\end{equation}
where, $\psi_k$ is the cost at step $k$ and $V_{k+1}$ is cost-to-go from time $k+1$ to $N$. The term $\psi_k$ is defined based on (\ref{eqn:general_stepcost}) as
\begin{equation}\label{eqn:cost_step}
{{\psi }_{k}}={{\left( {{x}_{k}}-x_{k}^{d} \right)}^{T}}{{Q}_{k}}\left( {{x}_{k}}-x_{k}^{d} \right)+u_{k}^{T}{{R}_{k}}{{u}_{k}}
\end{equation}
If we find the optimal control, state and cost for all values from $k+1$ to $N$, the optimal value for the step $k$ to $k+1$ can be found out using the functional equation of dynamic programming \cite{naidu2002optimal}, which is given by
\begin{align}\label{eqn:value}
V_{k}^{*}\left( {{x}_{k},\hat{x}_k} \right)=&\underset{_{u_k}}{\mathop{\min }}\,\big\{\mathop{\mathbb{E}}\left[\psi_k\right]+\mathop{\mathbb{E}}\left[ V_{k+1}^{*}\left( {{x}_{k+1},\hat{x}_{k+1}} \right)|{{x}_{k}},\hat{x}_k,{{u}_{k}} \right] \big\}
\end{align} 
For this problem considering cost per step as given in (\ref{eqn:cost_step}), the relationship in (\ref{eqn:value}) becomes
\begin{align}
\label{eqn:value_exp}
V_k^*(x_k,\hat{x}_k) = \min_{u_k} \big\{&(x_k-x_k^d)^TQ_k(x_k-x_k^d)+u_k^TR_ku_k \nonumber\\ 
&+\mathbb{E}[ V_{k+1}^{*}(x_{k+1},\hat{x}_{k+1}|x_k,\hat{x}_k,u^*_k)]\big\}
\end{align}
The optimal control $u^*_k$ can be obtained by solving (\ref{eqn:value_exp}) by using a value approximation method as described in \cite{todorov2005stochastic,lewis2012optimal}. Given a problem with value function dependent on the state $x_k$ and the estimate of the state $\hat{x}_k$, the approximate form of value under optimal control policy for achieving tracking of a desired state can be written as
\begin{align}\label{eqn:val_app}
V_k = x_k^TW_k^xx_k-2x_k^TW_k^r+e_k^TW_k^ee_k+W_k
\end{align}
where, ${{e}_{k}}\triangleq{{x}_{k}}-{{{\hat{x}}}_{k}}$ and $W_k^x$, $W_k^e$, $W_k^r$ and $W_k$ are the weightages. The variable ${{e}_{k}}$ is the estimation error between the current state of the system and the estimated state predicted by the forward model. The approximated form of value in (\ref{eqn:val_app}) is used for expanding the right hand side of expression in (\ref{eqn:value_exp}) to get
\begin{align}
\label{eqn:val_subs}
{{V}_{k}}= & (x_k-x_k^d)^TQ_k(x_k-x_k^d)+u_k^TR_ku_k \nonumber \\ 
& +\mathop{\mathbb{E}}[x_{k+1}^TW_{k+1}^xx_{k+1}+e_{k+1}^TW_{k+1}^ee_{k+1}\nonumber \\ 
& -2x_{k+1}^TW_{k+1}^r+W_{k+1}]
\end{align}
It is known that for any random variable ${{z}_{k}}$, $\mathop{\mathbb{E}}[z_k^TAz_k]$ can be expanded in terms of expectation ($\mathop{\mathbb{E}}$) and variance ($Var$) of $z_k$ as
\begin{align}
\label{eqn:prop}
\mathop{\mathbb{E}}[z_k^TAz_k]=\mathop{\mathbb{E}}[z_k]^TA\mathop{\mathbb{E}}[z_k]+Tr[ A(Var[z_k])]
\end{align}
where $Tr$ represents the trace operator. Using this property, the quadratic term in $x_{k+1}$ in (\ref{eqn:val_subs}) can be expanded as follows 
\begin{equation*}
\mathop{\mathbb{E}}[x_{k+1}^TW_{k+1}^xx_{k+1}]=\mathop{\mathbb{E}}[x_{k+1}]^TW_{k+1}^x\mathop{\mathbb{E}}[x_{k+1}]+Tr[ W_{k+1}^xVar[ x_{k+1}]]
\end{equation*}
Further, $\mathop{\mathbb{E}}\left[x_{k+1}\right]$ and $Var\left[x_{k+1}\right]$ is got from the state dynamics equation given by (\ref{eqn:stocsd}) and the following expression is obtained
\begin{align}
\label{eqn:expan_x1}
\mathop{\mathbb{E}}[x_{k+1}^TW_{k+1}^xx_{k+1}]= & (Ax_k+Bu_k)^TW_{k+1}^x(Ax_k+Bu_k) \nonumber\\ 
&+Tr[\alpha u_kB^TW_{k+1}^xBu_k\alpha]
\end{align}
The same property as given by (\ref{eqn:prop}) is used to expand the expectation of the quadratic term in $e_{k+1}$ in (\ref{eqn:val_subs}) and the expression is obtained as
\begin{align*}
\mathop{\mathbb{E}}[e_{k+1}^TW_{k+1}^ee_{k+1}]= &\mathop{\mathbb{E}}[x_{k+1}-\hat{x}_{k+1}]^TW_{k+1}^e\mathop{\mathbb{E}}[x_{k+1}-\hat{x}_{k+1}]\\
&+Tr\left[ W_{k+1}^eVar[x_{k+1}-\hat{x}_{k+1}]\right]
\end{align*}
Using the state dynamics given by (\ref{eqn:stocsd}) and forward model dynamics given by (\ref{eqn:FWD}) this expression can be written as
\begin{align}
\label{eqn:expan_e1}
\mathop{\mathbb{E}}[e_{k+1}^TW_{k+1}^ee_{k+1}]= & e_k^TA^TW_{k+1}^eAe_k \nonumber\\ 
&+Tr[\alpha u_kB^TW_{k+1}^eBu_k\alpha]
\end{align}
Also, $\mathop{\mathbb{E}}[x_{k+1}^TW_{k+1}^r]$ can be written as
\begin{equation}
\label{eqn:expan_r}
\mathop{\mathbb{E}}[x_{k+1}^TW_{k+1}^r]= {(W_{k+1}^r)}^T\mathop{\mathbb{E}}[x_{k+1}]={(W_{k+1}^r)}^T\left(Ax_k+Bu_k\right)
\end{equation}
Thus, the value function in (\ref{eqn:val_subs}) can be expanded using (\ref{eqn:expan_x1}), (\ref{eqn:expan_e1}) and (\ref{eqn:expan_r}) to get
\begin{align}
\label{eqn:value_new}
V_k= & (x_k-x_k^d)^TQ_k(x_k-x_k^d)+u_k^TR_ku_k \nonumber\\ 
& +(Ax_k+Bu_k)^TW_{k+1}^x(Ax_k+Bu_k)\nonumber\\ 
& +Tr[\alpha u_kB^TW_{k+1}^xBu_k\alpha] \nonumber\\ 
& +e_k^TA^TW_{k+1}^eAe_k+Tr[ cu_kB^TW_{k+1}^eBu_kc] \nonumber\\
& -2(W_{k+1}^r)^T(Ax_k+Bu_k)+W_{k+1}
\end{align}
Applying the necessary condition of optimality given by
\begin{equation}
\label{eqn:nec_1}
\left. \frac{\partial {V_k}}{\partial {u_k}} \right|_{u=u_k^*}=0
\end{equation}
with $V_k$ as in (\ref{eqn:value_new}), the optimal control signal $u^*_k$ at step $k$ can be obtained. Equation (\ref{eqn:nec_1}) leads to the following expression
\begin{align}
\label{eqn:nec_condn}
& R_ku^*_k + B^TW_{k+1}^xBu^*_k+Tr[\alpha B^TW_{k+1}^xB\alpha]u^*_k \nonumber\\ 
&+Tr[ \alpha B^TW_{k+1}^eB\alpha]u^*_k+B^TW_{k+1}^xAx_k-(W_{k+1}^r)^TB=0
\end{align}
which is grouped and rearranged to obtain the optimal control $u_{k}^{*}$ as
\begin{equation}\label{eqn:optcntrl}
u_{k}^{*}=-{{G}_{k}}{{x}_{k}}+{{b}_{k}}
\end{equation}
with the feedback gain given by
\begin{equation}\label{eqn:statefbc}
{{G}_{k}} =L_{k}^{-1}{{B}^{T}}W_{k+1}^{x}A
\end{equation} 
where,
\begin{align}
\label{eqn: gain}
&{{L}_{k}}\triangleq({{R}_{k}}+C_{k+1}^{x}+C_{k+1}^{e}+{{B}^{T}}W_{k+1}^{x}B)\\
& C_{k+1}^{x}\triangleq Tr\left( \alpha{{B}^{T}}W_{k+1}^{x}B\alpha \right), C_{k+1}^{e}\triangleq Tr\left( \alpha{{B}^{T}}W_{k+1}^{e}B\alpha \right)
\end{align} 
and the bias term is given by
\begin{equation}\label{eqn:bias}
{{b}_{k}}\triangleq L_{k}^{-1}(W{{_{k+1}^{r}})^{T}}B
\end{equation} 
Since there is no external state feedback information, the forward model is used to estimate the expectation of the current state of the system as in (\ref{eqn:FWD}), and the control signal is calculated on this estimated state as $u_k^*=-G_k\hat{x}_k+b_k$. In the next (Section {\ref{sec:OC_VA}), it is shown that inspite of calculating optimal control signal from estimate of the state, the corresponding optimal value function obtained takes the same optimal form as assumed in (\ref{eqn:val_app}). 

\par The update equations for the weight matrices $W_{k}^{x}$, $W_{k}^{e}$, $W_{k}^{r}$ and $W_k$ can be obtained by substituting the optimal control expression $-G_k\hat{x}_k+b_k$ in the optimal value function obtained in (\ref{eqn:value_new}). After simplification, it is observed that the the left out terms can be regrouped into terms quadratic in $x_k$, quadratic in $e_k$, linear in $x_k$ and constant term. By comparing these terms with similar terms in approximate value form in (\ref{eqn:val_app}), the weight update equations are obtained. The update equations turn out to be backward recursive equations dependent on system matrices $A$ and $B$, the feedback gain $G_k$ and the bias $b_k$. They are given by
\begin{align}
\label{eqn:weights}
W_{k}^{x} &={{Q}_{k}}+{{A}^{T}}W_{k+1}^{x}A-{{A}^{T}}W_{k+1}^{x}B{{L}^{-1}}{{B}^{T}}W_{k+1}^{x}A \nonumber\\ 
W_{k}^{e} &={{A}^{T}}W_{k+1}^{e}A+{{A}^{T}}W_{k+1}^{x}B{{L_k}^{-1}}{{B}^{T}}W_{k+1}^{x}A \nonumber\\ 
W_{k}^{r} &=AW_{k+1}^{r}+{{Q}_{k}}x_{k}^{d}-A^TW_{k+1}^xBL_k^{-1}W{{_{k+1}^r}^T}B \nonumber \\ 
W_{k} &={{W}_{k+1}}+2x{{_{k}^{d}}^{T}}{{Q}_{k}}x_{k}^{d}-B^TW_{k+1}^rL_k^{-1}W{{_{k+1}^r}^T}B
\end{align}
Note that in (\ref{eqn:weights}), $G_k$ and $b_k$ are written in terms of their expansions given by (\ref{eqn:statefbc}) and (\ref{eqn:bias}), respectively. The weight matrices at final step $N$ can be obtained as $W_{N}^{x}=Q_N\text{, }W_{N}^{r}={{Q}_{N}}x_{N}^{d}\text{, }W_{N}^{e}=0$\text{ and } $W_N=2{x_N^d}^{T}Q_Nx^d_{N}$, by substituting all weights in step $N+1$ as zero in (\ref{eqn:weights}). Subsequently, these weights are used in (\ref{eqn:weights}) to calculate the weights in the second last step $N-1$ and recursively backward till first step. Using these weights, the feedback control gain ${{G}_{k}}$ and the bias ${{b}_{k}}$ are calculated backward from the last time step using the (\ref{eqn:statefbc}) and (\ref{eqn:bias}) respectively. The gains ${{G}_{k}}$ and ${{b}_{k}}$ are used to calculate the optimal control signal for each step using $-G_k\hat{x}_k+b_k$, while simultaneously propagating the forward model to get an estimate of the state $\hat{x}_k$.
\subsection{Verification of optimality of value function}\label{sec:OC_VA}
\par The optimal control signal obtained by solving the necessary conditions of optimality was in terms of the actual state $x_k$. However, since the state feedback is not available, the estimate of the state $\hat{x}_k$ given by the forward model is used for calculating the optimal control. This necessitates to check whether the new optimal value function obtained based on this optimal control calculated from $-G_k\hat{x}_k+b_k$ will be of the same form as assumed in (\ref{eqn:val_app}). To carry out this check, the optimal control expression with $x_k$ replaced by $\hat{x}_k$ is substituted in (\ref{eqn:value_new}) and the optimal value function is obtained. The optimal value after regrouping the terms quadratic in $u^*_k$ is given by,
\begin{align}
\label{eqn: opt_val}
V_k= & {(x_k-x_k^d)}^TQ_k(x_k-x_k^d)\nonumber\\
&+\big{\{}(-{{G}_{k}}{{\hat{x}}_{k}}+{{b}_{k}})^T(R_k+B^TW_{k+1}^xB+C^x_{k+1}+C^e_{k+1})\nonumber\\
&\hspace{0.4cm}(-{{G}_{k}}{{\hat{x}}_{k}}+{{b}_{k}})\big{\}}+(Ax_k)^TW_{k+1}^x(Ax_k)\nonumber\\
&+e_k^TA^TW_{k+1}^eAe_k-2W{{_{k+1}^{r}}^{T}}(Ax_k+Bu^*_k)+W_{k+1}
\end{align}
This function is expanded using (\ref{eqn:statefbc}) and (\ref{eqn:bias}) and obtained as
\begin{align}
\label{eqn: opt_valgrpd}
V_k= & x_k^TQ_kx_k+{x_{k}^d}^TQ_kx_{k}^d-2x_k^TQ_kx_k^d+\hat{x}_k^TZ\hat{x}_k \nonumber\\
& -2B^TW_{k+1}^xA\hat{x}_kL_k^{-1}W{{_{k+1}^r}^T}B+B^TW_{k+1}^{r}{L_k}^{-1}W{{_{k+1}^r}^T}B \nonumber\\
& +x_k^TA^TW_{k+1}^{x}Ax_k -2{\hat{x}_k}^TZx_k +2x_k^TA^TW_{k+1}^xBL_k^{-1}W{{_{k+1}^r}^T}B \nonumber\\
& +e_k^TA^TW_{k+1}^eAe_k-2W{{_{k+1}^r}^T}Ax_k \nonumber\\
&+2W{{_{k+1}^r}^T}BL_k^{-1}B^TW_{k+1}^xA\hat{x}_k \nonumber\\
& -2B^TW{{_{k+1}^r}}L_k^{-1}W{{_{k+1}^r}^T}B +W_{k+1}
\end{align}
The expansion consists of terms which are quadratic and linear in $x_k$, quadratic in $e_k$, quadratic and linear in $\hat{x}_k$ as well as constants. In (\ref{eqn: opt_valgrpd}), the fifth term and twelfth term cancel each other. The terms in $\hat{x}_k$ can be eliminated by replacing the fourth term and eighth term together using the expression
\begin{align}
\label{eqn: rep_xhat}
\hat{x}_k^TZ\hat{x}_k-2{\hat{x}_k}^TZx_k = (x_k-\hat{x}_k)^TZ(x_k-\hat{x}_k)-x_k^TZx_k
\end{align}
where, $Z=A^TW_{k+1}^xBL_k^{-1}B^TW_{k+1}^xA$ and the first term is quadratic in $e_k$ while the second term is quadratic in $x_k$. The reduced optimal value function equation then becomes
\begin{align}
\label{eqn: opt_valred}
V_k=& x_k^T(Q_k+{{A}^{T}}W_{k+1}^{x}A-Z)x_k+e_k^T(A^TW_{k+1}^eA+Z)e_k \nonumber\\
& -2x_k^T\big{(}W{{_{k+1}^r}^T}A+Q_kx_k^d-A^TW_{k+1}^xBL_k^{-1}W_{k+1}^rB\big{)} \nonumber\\
& -B^TW{{_{k+1}^r}^T}L_k^{-1}W_{k+1}^{r}B +{x_{k}^d}^TQ_kx_{k}^d+W_{k+1}
\end{align}
It is clear from (\ref{eqn: opt_valred}) that eventhough a modified optimal control signal based on the estimate of the state is used in place of the actual expression obtained in (\ref{eqn:optcntrl}), the optimal value function remains of the same optimal form which was assumed in (\ref{eqn:val_app}). Hence, the assumption of using estimated of the state in place of actual state is justified and the approximated value of the form given in (\ref{eqn:val_app}) can be used.
\par In this section, we obtained a desired velocity tracking optimal controller which converts the estimate of the actual state (which is given by the forward model in the internal feedback loop) to produce the control signals that would serve as motor commands to the oculomotor plant. The mean trajectories of displacement and velocity are simulated by substituting optimal control signal in the expectation of the state dynamics equation given by (\ref{eqn:stocsd}). To verify the proposed model of saccade control based on velocity tracking, the predictions of this saccade generation framework are tested using behavioral data of saccades obtained from the human experiment. The details of the experiment are discussed in the next section.\\
\section{Experimental Data}\label{sec:exp}
\par The behavioral data were collected from 20 healthy subjects in the age group of $20-33$ years in a laboratory set-up. Each of the 20 participants carried out repeated trials of saccades to the same targets during the experiment. The participants consisted of an equal number of men and women. Participants were monetarily compensated for taking part in the experiment. Informed consent was obtained from all participants in accordance with the guidelines of the Institute Human Ethics Committee of the Indian Institute of Science (IHEC No: 1-28102016 approved on 28-10-2016).
\subsection{Experiment Setup and Design}\label{sec:design}

\par Saccadic eye movement data used in this study was collected using an IR based pupil tracking camera (ISCAN, Boston, USA) at a sampling rate of $240$ Hz. The setup is shown in Fig. \ref{fig:setup}. 
\begin{figure}[!h]
\centering
\includegraphics[width=0.45\textwidth,height=0.45\textwidth]{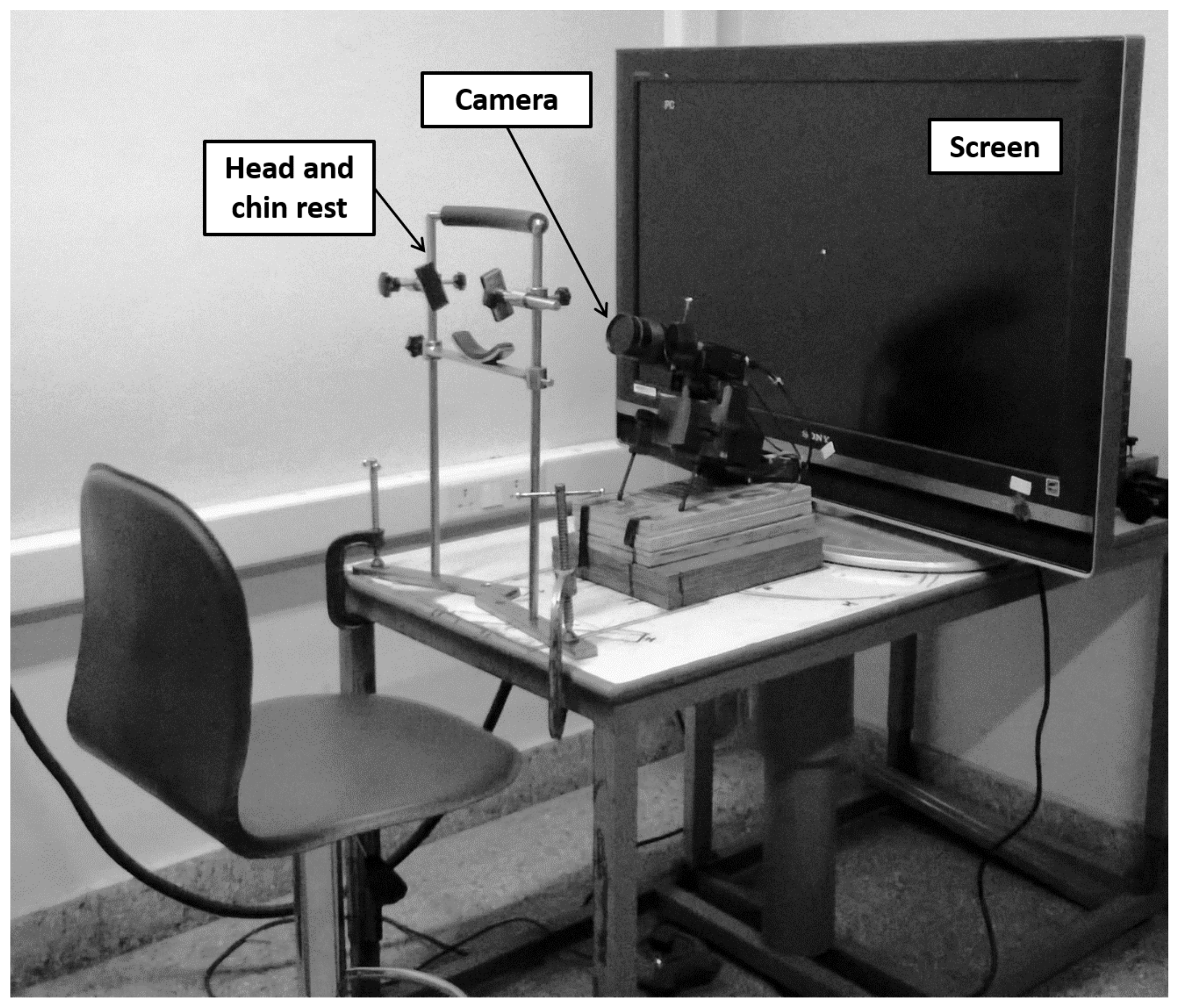}
\caption{Experimental setup for saccade data collection}
\label{fig:setup}
\end{figure}
Each participant was made to sit in front of the LCD screen on which stimuli were shown. They were also made to rest their chin such that the eye-level matched the center of the screen. At the same time, the head was locked at the temple to obtain eye movements exclusively in the absence of any head movements.
\begin{figure}[!h]
\centering
\includegraphics[width=0.45\textwidth,height=0.45\textwidth]{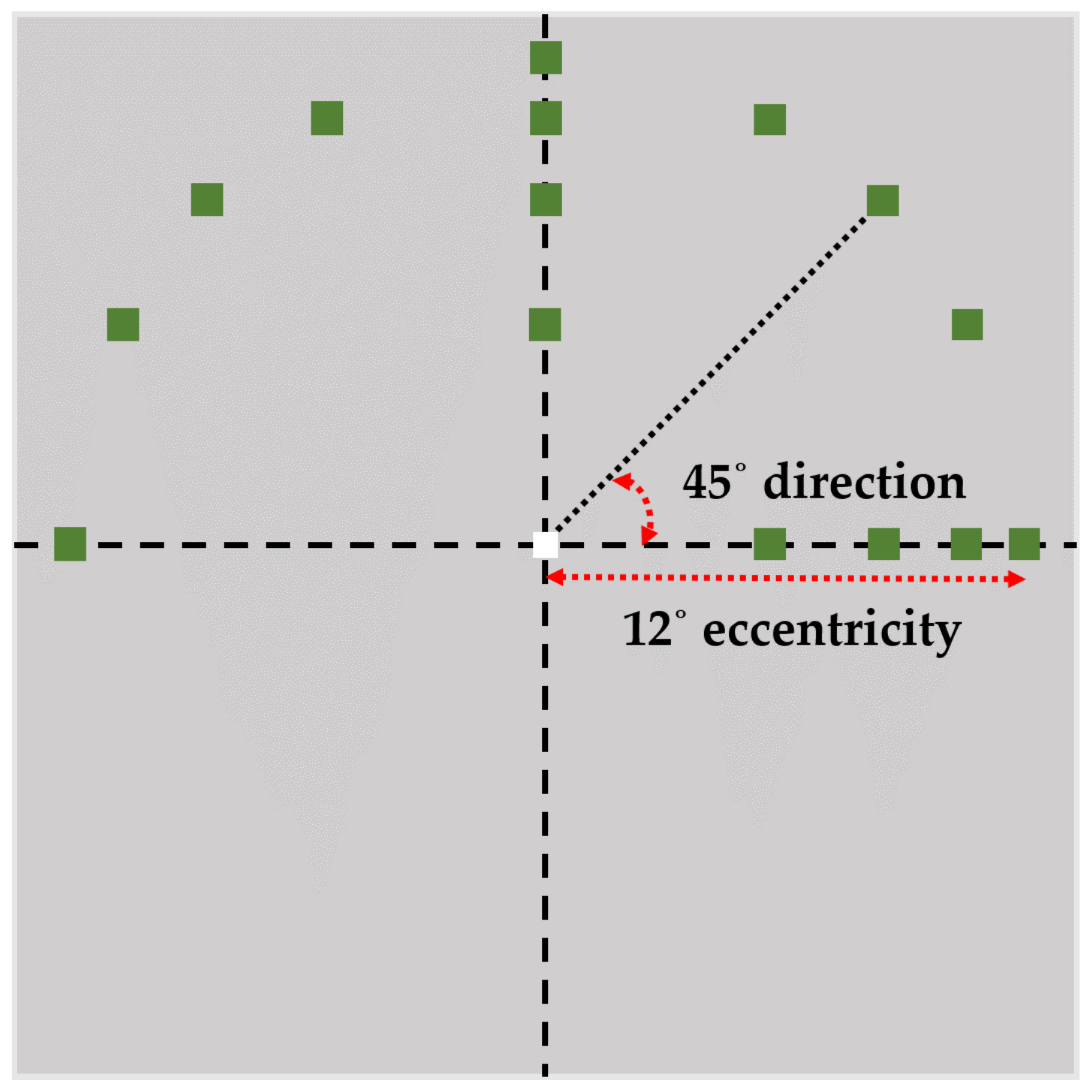}
\caption{Task paradigm showing all stimulus positions}
\label{fig:paradigm}
\end{figure}
\par The experiment had 15 different targets shown on the upper half of the screen in a randomized fashion as shown in Fig. \ref{fig:paradigm}. It included targets in 9 different directions including $0^\circ, 30^\circ, 45^\circ, 60^\circ, 90^\circ, 120^\circ, 135^\circ, 150^\circ$ and $180^\circ$ at an eccentricity of $12^\circ$ as well as targets at three different eccentricities of $6^\circ, 8.5^\circ$ and $10.4^\circ$ in both $0^\circ$ and $90^\circ$ directions. Each trial started with a white fixation point shown in the middle of the screen. After a $400\pm60$ ms delay, the fixation box disappeared and a green target appeared at the periphery. The subject was instructed to make one single saccadic eye movement to the target as soon as it appeared. A green tick mark on the screen and a beep sound gave feedback to the subject about the success of the trial.
\subsection{Data Analysis}\label{sec:anal}
\par Saccade displacement was calculated from horizontal and vertical positions obtained from the eye camera. It was fit using a polynomial function of seventh order and differentiated to obtain the velocity profiles. The saccade start and end were detected based on a velocity criterion. The time at which the velocity of the eye exceeds $10\%$ of the peak velocity was marked as the start of the saccade and the time point where the velocity reduces below the same threshold after the peak velocity was considered the end of the saccade. Since each trial produced saccades of different duration, the time was normalized into equal bins for the entire saccade duration. This allowed the calculation of the mean saccade trajectory across trials at these bins.\\


\section{Model parameter estimation}\label{sec:est}
\par The model has two free parameters, $q$ and $c$, which are tuned individually for each subject. The model parameters were estimated by a two-stage process. The parameter $q$ in the error weightage matrix in the cost function (refer Section \ref{sec:OC}) was estimated such that the error between the velocity prediction of the model and experimental velocity was minimized. This was done after setting the noise scaling parameter $\alpha$ to zero. The error is defined as
\begin{equation}\label{eqn:err}
e=\frac{\sqrt{\sum^{m}_{j=1}[{z}_{\mu}^{expt}(t_j)-{z}_{\mu}^{pred}(t_j)]^2}}{\sqrt{\sum^{m}_{j=1}[{z}_{\mu}^{expt}(t_j)]^{2}}}.\\
\end{equation}
where, $z$ is defined as the variable of interest. It was taken as the angular velocity ($\dot{\theta}$) for obtaining the free parameters. Here, $\dot{\theta}_{\mu}^{expt}$ is defined as the mean angular velocity obtained for individual subjects based on experimental data and $\dot{\theta}_{\mu}^{pred}$ is the model's prediction of the same. The second stage in the estimation process was of tuning the noise scaling parameter $\alpha$ (refer Section \ref{sec:noise}). This parameter was obtained by minimizing the same error as described in (\ref{eqn:err}), but with the value of $q$ in the model fixed at the estimated value. The parameter ranges and guess values for the optimization are obtained by scanning the parameter space. The parameters were estimated using saccades from $12^\circ$ eccentricity in $180^\circ$ direction (horizontal leftward saccades). These were used for all further predictions of the model. The errors in fit were also quantified using (\ref{eqn:err}) by defining the variable $z$ as ${\theta}$ or $\dot{\theta}$ depending on whether fit error in displacement was quantified or fit error in velocity, respectively. The errors are all expressed in percentages.\\


\section{Results Analysis}\label{sec:results}
\par An example of saccade trajectories made by a subject across multiple trials for different oblique targets is shown in \ref{fig:raw_data}. Further analyses were carried out on the mean of the trajectories calculated across trials, which is shown with black dashed lines in the figure. 
\begin{figure}[!htbp]
\centering
\includegraphics[width=0.4\textwidth,height=0.4\textwidth]{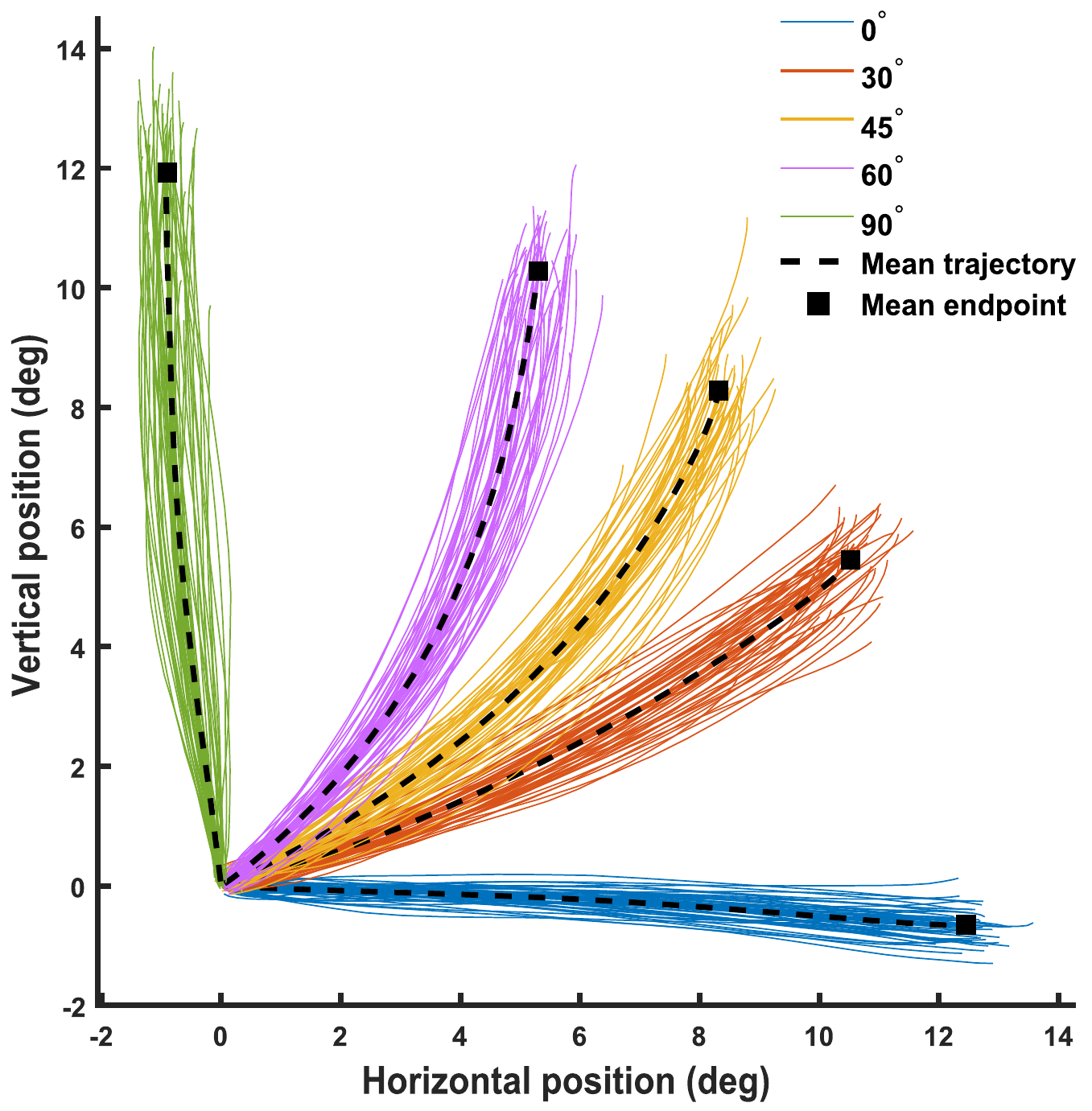}
\caption{Example of saccades made to different target directions by a subject}
\label{fig:raw_data}
\end{figure}
The mean angular displacement and angular velocity for an example subject who participated in the experiment are shown in Fig. \ref{fig:mean_kin} for four different targets amplitudes of $6^\circ, 8.5^\circ,10.4^\circ$ and $12^\circ$. The mean angular displacement is typically a monotonously increasing function of time while the mean angular velocity peaks towards the middle of the movement and then decreases. Further, saccadic eye movements with larger amplitudes have higher peak velocities as is observed in general \cite{bahill1975main,harris2006main}.
\begin{figure}[!htbp]
\centering
\includegraphics[width=0.49\textwidth,height=0.32\textwidth]{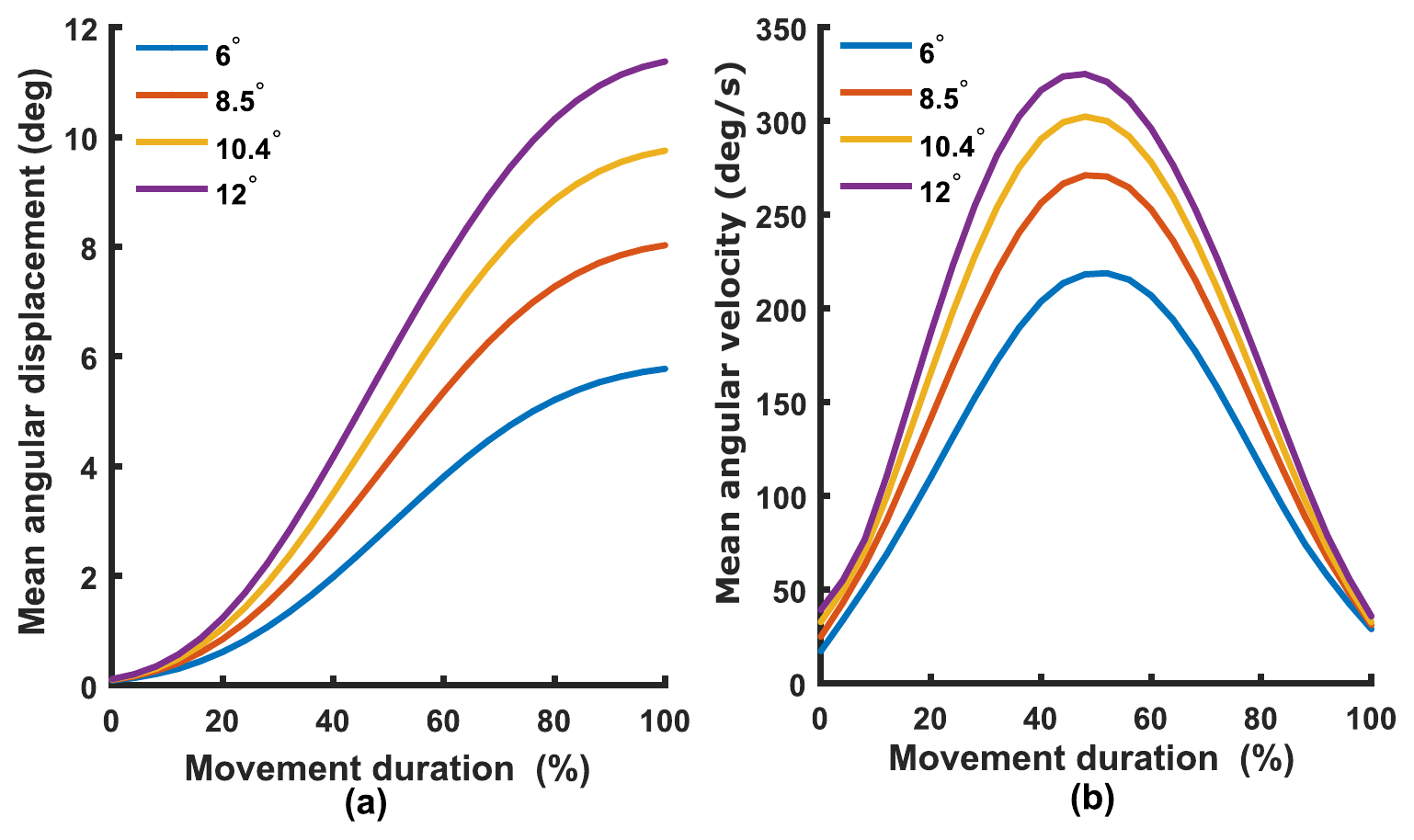}
\caption{Mean angular displacement and velocity profiles of saccades to different target amplitudes for an example subject}
\label{fig:mean_kin}
\end{figure}
\par The proposed model's free parameters were estimated from $12^\circ$ horizontal leftward saccades as described in Section \ref{sec:est} using the experimental mean angular velocity. An example fit for an individual subject is shown in Fig. \ref{fig:vel_fit}-(a). The errors in the fit as given by (\ref{eqn:err}) and is shown for all the 20 subjects in Fig. \ref{fig:vel_fit}-(b). The mean error in velocity fit was $0.65\%$ and the standard deviation was $0.13\%$. The fit errors for all subjects were less than $1\%$. 
\begin{figure}[!htbp]
\centering
\includegraphics[width=0.45 \textwidth]{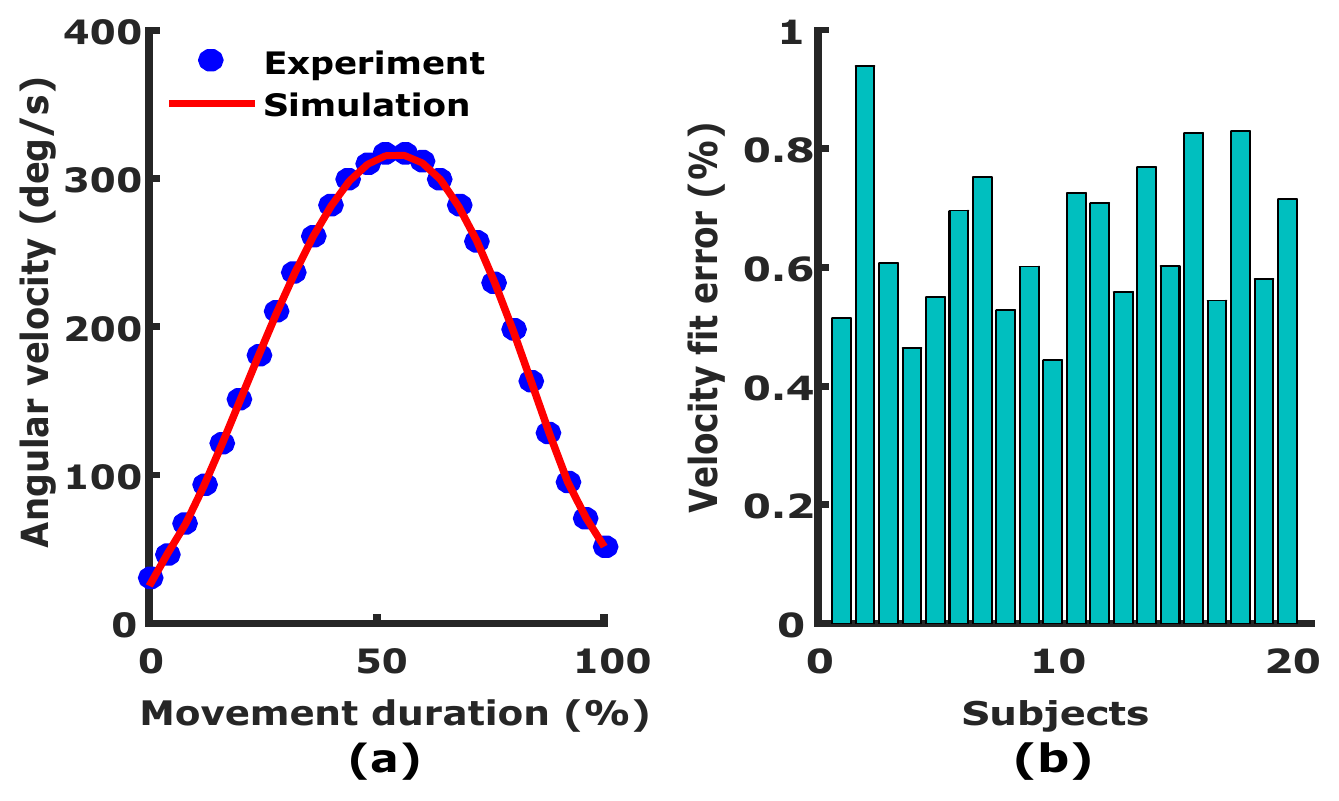}
\caption{Angular velocity fit of the proposed model.}
\label{fig:vel_fit}
\end{figure}
The predictions of mean angular displacement for the same has been presented separately to validate the model. The prediction error was also calculated and expressed in $\%$ by using (\ref{eqn:err}) as described in Section \ref{sec:est}. Fig. \ref{fig:disp_pred}-(a) shows an example of the prediction of mean angular displacement for the same $12^\circ$ horizontal leftward saccades which is used for fitting the parameters. The prediction errors quantified for all 20 subjects are shown in Fig. \ref{fig:disp_pred}-(b). All prediction errors of the model were less than $5\%$. The mean error in prediction of angular displacement, in this case, was $3.2\%$ with a standard deviation of $0.5\%$. Thus, the model can capture the saccade kinematics with low errors.
\begin{figure}[h]
\centering
\includegraphics[width=0.48\textwidth]{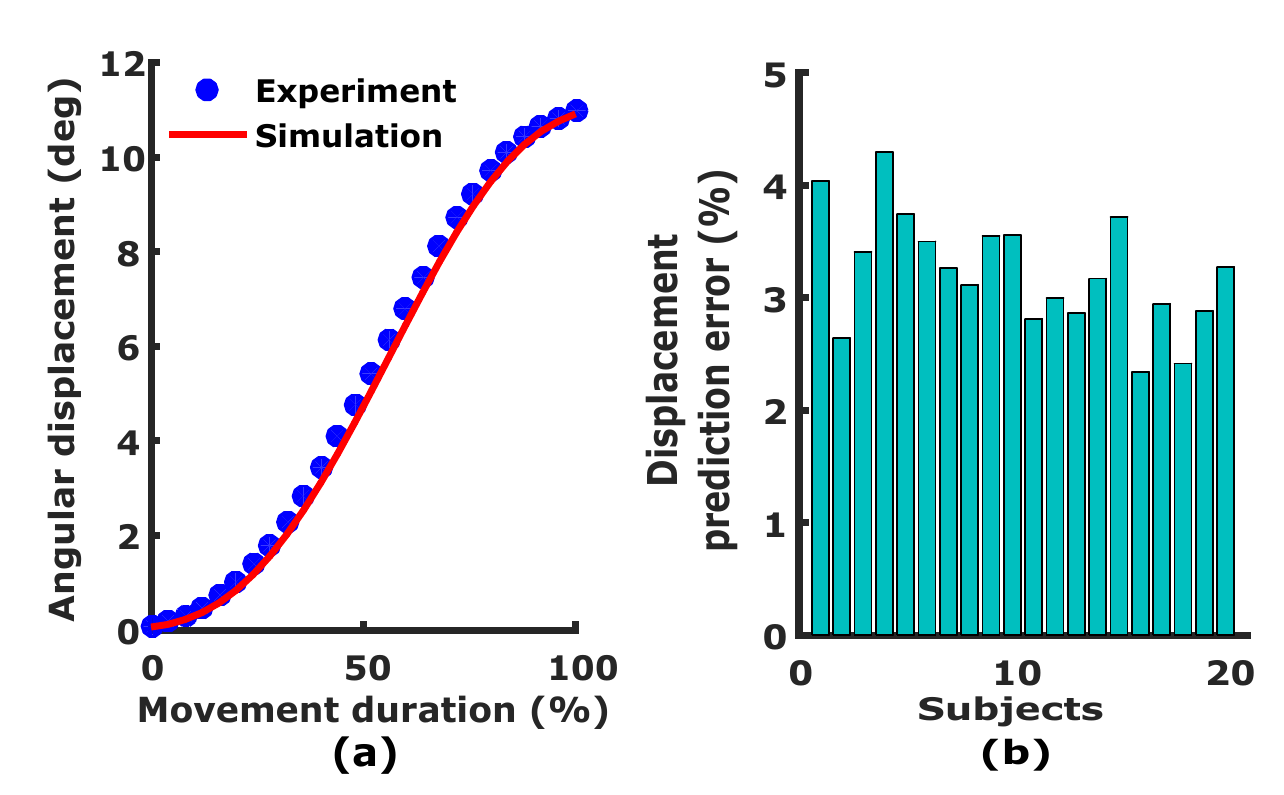}
\caption{Model prediction of mean angular displacement of saccades for $12^\circ$ horizontal leftward saccades.}
\label{fig:disp_pred}
\end{figure}
\par To check the generalizability of the model to other amplitudes, the model predictions of mean angular displacement and velocity of saccades made to the horizontal rightward targets at $6^\circ, 8.5^\circ,10.4^\circ$ and $12^\circ$ were simulated. The free parameters were set to the already estimated values from $12^\circ$ horizontal leftward saccades. The errors in prediction for all amplitudes across 20 subjects are quantified in Fig. \ref{fig:amp_err} for each of the amplitudes. The subplot Fig. \ref{fig:amp_err}-(a) shows the prediction errors for angular displacement while that in Fig. \ref{fig:amp_err}-(b) shows the prediction errors for angular velocity. Displacement prediction errors across subjects for all amplitudes were less than $6\%$, while errors for velocity were all less than $1\%$. In general, the errors in the prediction of angular velocity were much lesser compared to the prediction error in angular displacement for all target amplitudes. This is expected given that the model is tracking the desired velocity. The mean and standard deviation of the prediction errors across subjects are presented in the Table. \ref{table: amp_err}. 
\begin{figure}[!h]
\centering
\includegraphics[width=0.48\textwidth,height=0.45\textwidth]{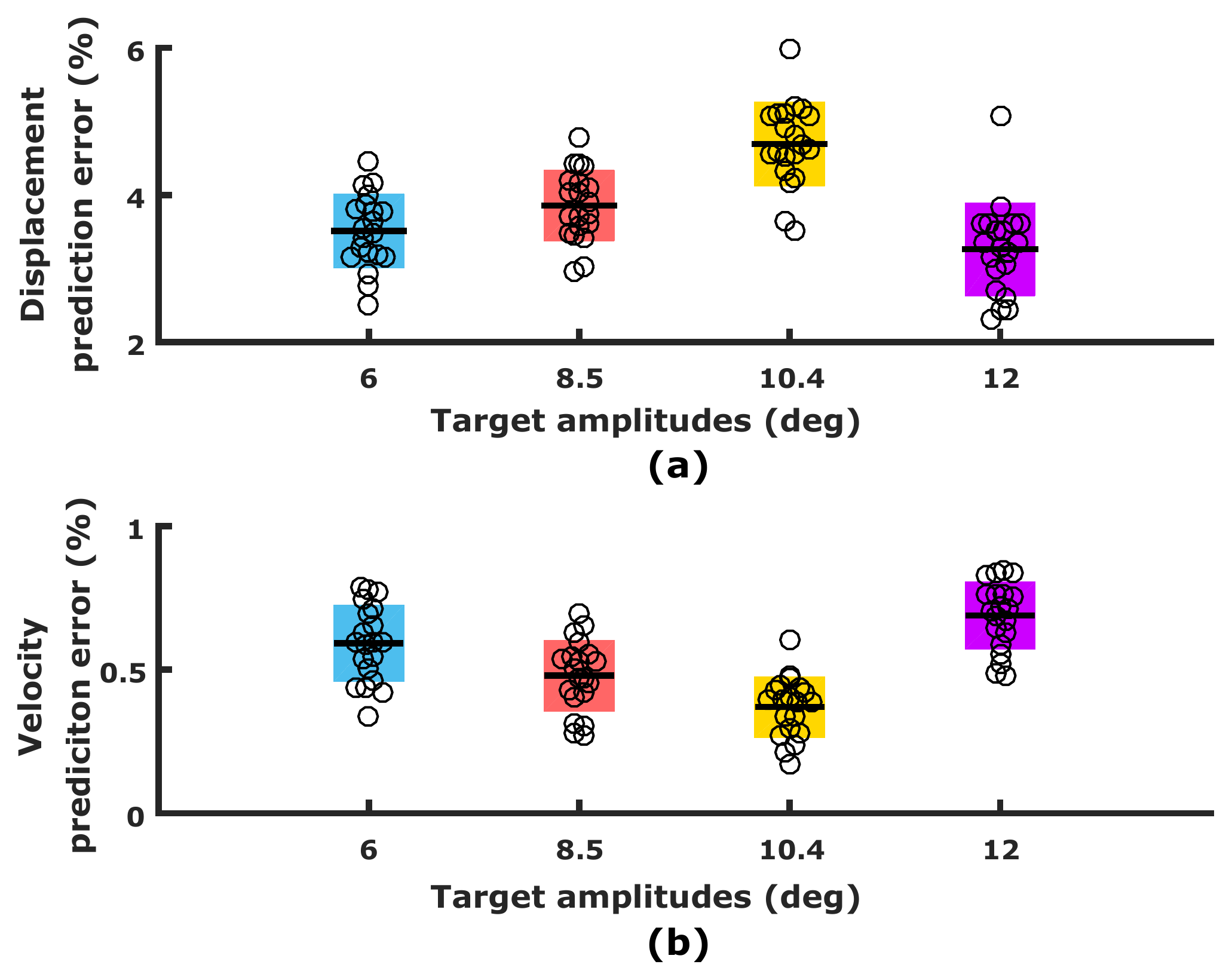}
\caption{Model prediction errors in angular displacement and velocity for saccades to different target amplitudes.}
\label{fig:amp_err}
\end{figure}
\begin{table} [!htbp]
\centering
\caption {Model prediction errors across amplitudes}
\label{table: amp_err}
\begin{tabular}{l c c}
\toprule
Target amplitude &\multicolumn{2}{c}{Prediction error (mean$\pm$std\textsuperscript{*}\%)}\\
\midrule
{} &Displacement &Velocity\\
$6^\circ$ &$3.51\pm0.49$ &$0.59\pm0.13$\\
$8.5^\circ$ &$3.85\pm0.47$ &$0.47\pm0.12$\\ 
$10.4^\circ$ &$4.69\pm0.56$ &$0.37\pm0.10$\\
$12^\circ$ &$3.26\pm0.62$ &$0.68\pm0.11$\\
\bottomrule
\multicolumn {3}{l}{*\footnotesize{standard deviation}}
\end{tabular}
\end{table}
\par Further validation of the model was also carried out by verifying the model's ability to predict the main sequence relationship between saccade amplitude and peak velocity (Fig. \ref{fig:mainseq}). The close correspondence between the fit and the data indicated that the model can capture the main sequence relationship between saccade amplitude and peak velocity (see example in Fig. \ref{fig:mainseq}-(a)). To quantify the same across subjects, the prediction errors in mean saccade amplitude and mean saccade peak velocity between the experimental data and model simulation were quantified and are shown in Fig. \ref{fig:mainseq}-(b). The errors were all less than $2.5\%$. However, the non-linearity of this relationship at higher amplitudes could not be tested as all the amplitudes available in the experimental data were below $12^\circ$, which was in the linear part of the main sequence.
\begin{figure}[!h]
\centering
\includegraphics[width=0.41\textwidth]{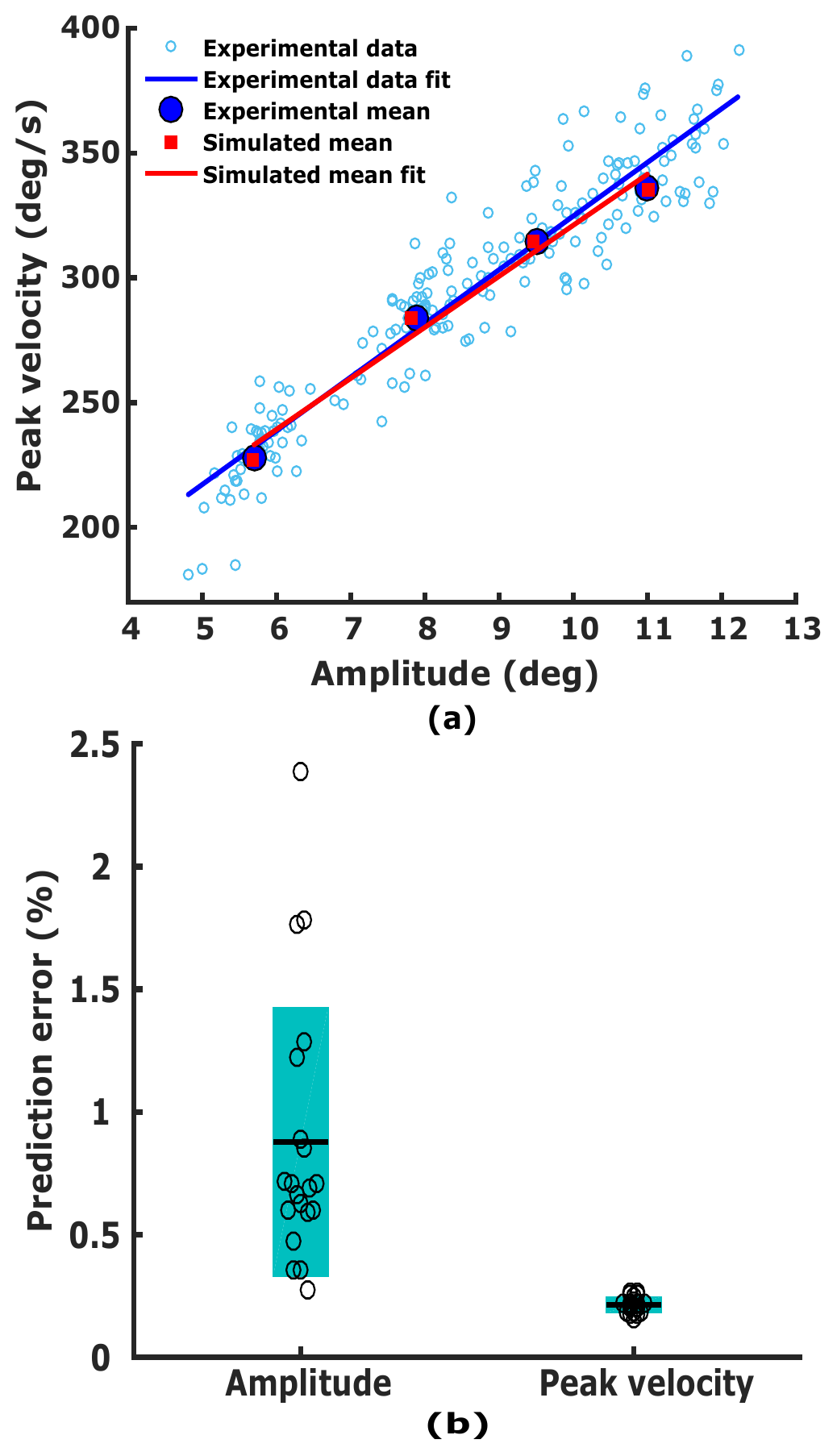}
\caption{Model prediction of main sequence relationship between saccade amplitude and peak velocity.}
\label{fig:mainseq}
\end{figure}
\par All the results presented until now were based on horizontal saccades. Hence, to further test the capability of the proposed velocity tracking model, simulations of saccades made to different directions were carried out. For this purpose, the modified model with system dynamics with system matrices as mentioned in (\ref{eqn: obldyn}) was used to generate saccades made to $12^\circ$ eccentricity in oblique directions of $30^\circ, 45^\circ, 60^\circ$ and vertical direction of $90^\circ$, relative to the horizontal axis. The same parameters estimated with $12^\circ$ leftward saccades were used. An example of the model's prediction for one subject for position and velocity are shown in Figs. \ref{fig:obl_full}-(a) and \ref{fig:obl_full}-(b), respectively. The $0^\circ$ target is the same as horizontal $12^\circ$ in the across target amplitude analysis and is presented again for completion. The quantification of the error in prediction for all 20 subjects is shown for all directions in Fig. \ref{fig:obl_full}-(c) for displacement and in Fig. \ref{fig:obl_full}-(d) for angular velocity.
\begin{figure}[h]
\centering
\includegraphics[width=0.5 \textwidth]{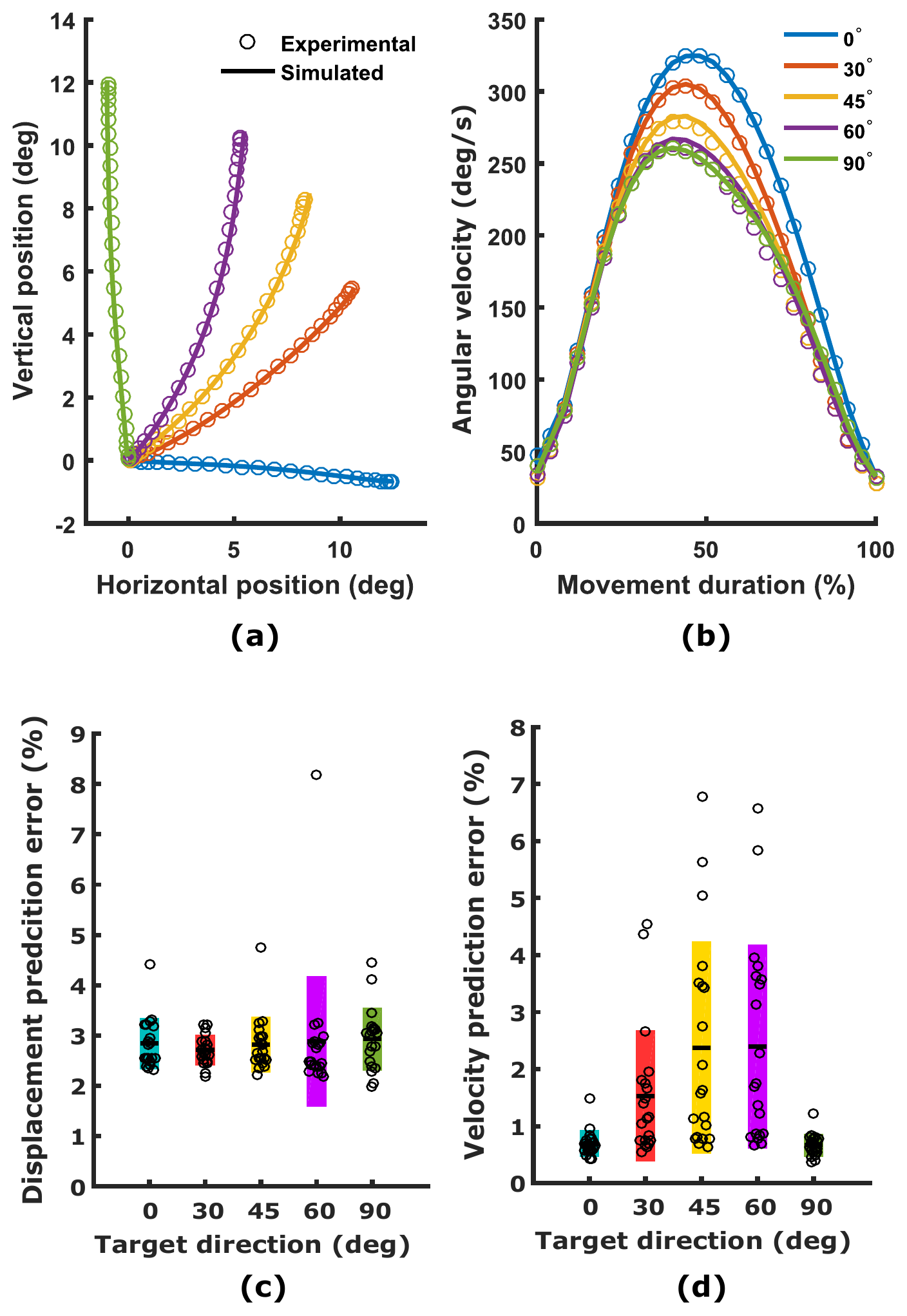}
\caption{Model prediction of saccade trajectory for oblique saccades and error quantification.}
\label{fig:obl_full}
\end{figure}
The mean and standard deviation in these predictions for oblique and vertical saccades are given in Table. \ref{table: dir_err}.
\begin{table} [!htbp]
\centering
\caption {Model prediction errors across directions}
\label{table: dir_err}
\begin{tabular}{l c c}
\toprule
Target direction &\multicolumn{2}{c}{Prediction error (mean$\pm$std\textsuperscript{*}\%)}\\
\midrule
{} &Displacement &Velocity\\
$0^\circ$ &$2.84\pm0.50$ &$0.69\pm0.22$\\
$30^\circ$ &$2.71\pm0.29$ &$1.53\pm1.14$\\ 
$45^\circ$ &$2.82\pm0.54$ &$2.37\pm1.85$\\
$60^\circ$ &$2.88\pm1.28$ &$2.39\pm1.77$\\
$90^\circ$ &$2.92\pm0.62$ &$0.66\pm0.19$\\
\bottomrule
\multicolumn {3}{l}{*\footnotesize{standard deviation}}
\end{tabular}
\end{table}
Although the prediction error in the case of oblique saccades is generally higher than those of the horizontal saccades especially in case of velocity, the predictions were still reasonably good. The errors on an average were $2.84\pm0.65\%$ for angular displacement and $1.53\pm1.03\%$ for angular velocity. This may be because, unlike horizontal saccades, oblique and vertical saccades involve more oculomotor muscles and there might be some interactions, at the oculomotor plant or before, which are not being considered in this work. Another possible reason could be that the parameters fixed based on horizontal saccades may not be optimal for the oblique and vertical saccades.\\


\section{Conclusion}\label{sec:conc}
\par A new velocity tracking optimal feedback control model has been proposed for saccade generation. The model also incorporated stochasticity in the saccadic system by considering an additive signal-dependent noise on the control signal that forms the input to the oculomotor plant. The model was able to predict the angular displacement profiles of horizontal saccades made to different amplitudes with high accuracy. It is also shown that the main sequence relationship between saccade amplitude and peak velocity was captured by the model in the linear range. Model performance is also validated for oblique saccades and vertical saccades made to targets in different directions. This model shows from a computational viewpoint how the brainstem saccade generation system could use a desired velocity signal at its input for producing accurate saccades. Thus, this study adds further evidence to the latest findings from neurophysiological studies that suggested input in the form of velocity being present in the saccadic system. 
\par Although the model does not disprove any of the existing models which are based only on end-point displacement; it suggests that the use of two different forms of desired inputs: (i) endpoint displacement, which is utilized as suggested by earlier models and (ii) velocity, which is proposed by this new model; may both be viable algorithms to make the saccadic system more robust. In this regard, this study provides another example of how redundancy may be a general principle of the motor system. How and when the system chooses to use one form of information over the other is an interesting question that needs further investigation.\\


\section*{Acknowledgments}
The experimental work was carried out in Visuomotor lab, Centre for Neuroscience, Indian Institute of Science (IISc) and was supported by grants from the Department of Biotechnology (DBT) - IISc partnership program. The authors are grateful to the participants for their time and effort. The authors would like to thank Dr. Kapil Sachan, former PhD student, and Niranjan Chakrabhavi, PhD student, IISc for their valuable suggestions in enhancing the paper presentation.\\

\bibliographystyle{IEEEtran}
\bibliography{References}


\end{document}